**Diamond photonics platform enabled by femtosecond laser writing**

Belén Sotillo[1], Vibhav Bharadwaj[1], J. P. Hadden[2], Masaaki Sakakura[4], Andrea Chiappini[3], Toney Teddy Fernandez[1], Stefano Longhi[1], Ottavia Jedrkiewicz[5], Yasuhiko Shimotsuma[4], Luigino Criante[6], Roberto Osellame[7], Gianluca Galzerano[7], Maurizio Ferrari[3], Kiyotaka Miura[4], Roberta Ramponi[7], Paul E. Barclay[2], Shane Michael Eaton[1],*

[1]Dipartimento di Fisica, Politecnico di Milano, Milano, Italy

[2]Institute for Quantum Science and Technology, University of Calgary, Calgary, Canada

[3]Istituto di Fotonica e Nanotecnologie (IFN)-CNR, CSMFO and FBK-CMM, Trento, Italy

[4]Office of Society-Academia Collaboration for Innovation and Department of Material Chemistry, Kyoto University, Kyoto, Japan.

[5]IFN-CNR and CNISM Udr Como, Como, Italy

[6]Center for Nano Science and Technology, Istituto Italiano di Tecnologia, Milano, Italy

[7]IFN – CNR, Milano, Italy

*shanemichael.eaton@polimi.it

**Abstract**

Diamond is a promising platform for sensing and quantum processing owing to the remarkable properties of the nitrogen-vacancy (NV) impurity. The electrons of the NV center, largely localized at the vacancy site, combine to form a spin triplet, which can be polarized with 532-nm laser light, even at room temperature. The NV's states are isolated from environmental perturbations making their spin coherence comparable to trapped ions. An important breakthrough would be in connecting, using waveguides, multiple diamond NVs together optically. However, still lacking is an efficient photonic fabrication method for diamond akin to the photolithographic methods that have revolutionized silicon photonics. Here, we report the first demonstration of three dimensional buried optical waveguides in diamond, inscribed by focused femtosecond high repetition rate laser pulses. Within the waveguides, high quality NV properties are observed, making them promising for integrated magnetometer or quantum information systems on a diamond chip.

**Introduction**

Usually admired for its beauty, diamond has other important qualities for technological applications: it is the hardest naturally occurring substance, offers high thermal conductivity and is transparent to ultraviolet light. Researchers recently discovered that diamond is an ideal substrate for spintronics and quantum information thanks to the properties of the nitrogen-vacancy (NV) center[1,2]. The negatively charged NV center impurity comprises a nitrogen atom adjacent to a vacancy. The electrons of the NV form a spin triplet, which can be polarized even at room temperature. One of the spin states fluoresces more brightly than the others which can be exploited for spin readout, with spin coherence times (~1 ms) comparable to trapped ions[3]. These properties make NVs attractive as a scalable platform for efficient sensing based on electron spins[4] and for quantum information systems[5].

Applications such as magnetic field sensing and coherent storage of light would benefit from the enhanced interaction provided by optical waveguides containing NVs[6]. Fabrication of waveguides in diamond to connect NVs is also compelling for long-range quantum entanglement[7]. However, it remains a significant challenge to fabricate optical waveguides in single-crystal diamond. To harness the potential of diamond NVs, a fabrication toolkit similar to the one that has revolutionized silica planar lightwave circuits[8] and silicon photonics[9] is desirable.

Optical waveguiding has been realized using a high refractive index material on a diamond substrate[10,11], but is limited by weak evanescent coupling to NVs near the surface. Single-crystal thin diamond films for waveguiding can be generated using the ion implantation assisted lift-off method[12], however the films suffer from residual stress which is detrimental for the fabrication of photonic structures, and NV centers within them often suffer from spectral diffusion due to charge traps on the surface[13]. Another approach uses heterogeneous diamond growth followed by oxygen plasma etching to create membranes to fabricate photonic crystals[14] and ring resonators[15,16], however it is difficult to remove tens of microns of diamond while maintaining smooth features needed for photonics applications. Using an angled plasma etching process, suspended triangular nanobeam waveguides were recently demonstrated[14,17] with losses < 10 dB/cm at visible wavelengths however for each design, a custom Faraday cage is required. Waveguides have also been demonstrated in diamond using isotropic etching[18] and ion beam writing[19] but as with other methods, optical circuits are restricted to 2D geometries.

We propose a disruptive technology using femtosecond laser writing to realize a 3D photonics toolkit for diamond. In this method, focused ultrashort pulses are nonlinearly absorbed in the bulk of a transparent material, leading to a localized modification[20]. In glasses this modification is a refractive index increase which enables 3D waveguide writing[20-28] but in crystals, the lattice is damaged yielding a decrease in refractive index[22]. Here, we laser-write closely spaced parallel lines in diamond to confine a guided optical mode between them. The high repetition rate of 500 kHz was found to reduce the formation of highly absorptive graphite, enabling optical waveguiding. Crucially, we find high quality NV center properties within the guiding region. Optical waveguides formed by directly focused femtosecond laser pulses can potentially be patterned in three dimensions to optically link or address NVs, to enable efficient excitation and collection, and make quantum information systems and spin-based sensing a reality.

**Results**

*The effect of repetition rate on bulk modification of diamond*

The initial focus of our study was to produce uniform modification lines below the surface of diamond, which could be used to form optical waveguides. At 500 kHz repetition rate, we found that an average power of 50 mW and a scan speed of 0.5 mm/s along <110> crystallographic directions produced uniform and reproducible modifications in the bulk of diamond. As shown in the optical microscope image in Fig. 1(a), the cross section of the laser-written track was approximately 5 μm transversely and 22 μm vertically. The significant vertical elongation of the line is attributed to the large spherical aberration caused by the mismatch between the index matching oil ($n = 1.5$) and diamond ($n = 2.4$). This asymmetry in the modification could be corrected using adaptive optics[29].

To better understand the structure of the femtosecond laser-written lines, μRaman spectroscopy was performed. As a reference, the pristine diamond[30] has a characteristic Raman peak centered at 1332 cm$^{-1}$ with a typical FWHM of around 2.3 cm$^{-1}$. For the above mentioned laser processing conditions, μRaman characterization (Fig. 1(b)) revealed that within the modification, there is a reduction of the intensity of the peak at 1332 cm$^{-1}$ to 15% of the original intensity along with an increase of its width by about 2 cm$^{-1}$, evidencing increased disorder in the diamond lattice. Moreover, at least two new bands appear: the G-peak at 1575 cm$^{-1}$ and the D-peak at 1360 cm$^{-1}$, showing a transformation of the $sp^3$ bonding of diamond into $sp^2$ bonding. In our case, the widths of D and G peaks (greater than 100 cm$^{-1}$) and the intensity ratio between them ($I$(D)/$I$(G) less and close to 1) indicate that these $sp^2$ clusters are mainly in an amorphous carbon phase rather than graphite[31]. This is in contrast to previous work with 1-kHz repetition rate Ti:Sapphire femtosecond lasers[32], which demonstrated the formation of micrographitic lines in diamond, which are not desirable for photonic devices due to the strong absorption.

In fact, when we reduce the repetition rate from 500 kHz to 5 kHz, we observe a sharper G-peak with a slight displacement to higher wavenumber (Fig. 1(c)), which implies a greater concentration of nanocrystalline graphite clusters[33]. We also found that the second order peaks at 2700 cm$^{-1}$ (2 D peak) and 2900 cm$^{-1}$ (D + G peak) appear for 5 and 25 kHz, but not for 500 kHz, evidencing increased graphitization at these lower repetition rates. Compared to the laser repetition rate, we found that the pulse energy and scan speed played a lesser role in the amount of graphitization.

*Bulk optical waveguiding in diamond*

Motivated by the μRaman results at 500 kHz repetition rate, we attempted to form waveguides in diamond using the type II method to produce an optical waveguide between two closely inscribed modification lines[22]. Figure 2(a) shows the transverse optical microscope image of the pair of laser-inscribed lines written 50 μm below the surface using the same laser processing conditions as the single line in Fig. 1(a). For 13-μm line separation, we produced for the first time to our knowledge, a buried optical waveguide in diamond using femtosecond laser writing. While the waveguide showed single mode behavior when scanning the input launch fiber transversely, when scanned vertically, modes could be excited at three different depths: the lowest loss mode was centered between the modification lines, as shown in Fig. 2(a). A mode could also be excited a few microns above and below this position (not shown in Fig. 2(a)). The mode field diameter (MFD) of the central and lowest loss mode is 10 μm × 11 μm at 635 nm wavelength with an insertion loss of 14 dB using butt-coupled single mode fibers (including a coupling loss of 2.8 dB/facet, Fresnel reflection loss of 0.3 dB/facet and propagation loss of 16 dB/cm). We also found optical waveguiding with similar losses in the near infrared (808 nm and 1550 nm). We attribute the propagation loss mainly to scattering loss due to the overlap of the optical mode with the modification tracks. The propagation loss would be much higher should graphite be present in the laser-written lines. By evaluating the attenuation of the optical mode[34] we estimate a waveguide damping loss of greater than $10^4$ dB/cm for type II modification tracks containing graphite. Further optimization of laser processing conditions to yield less overlap of the optical mode with the modification lines is expected to reduce the moderately high damping losses reported here.

At 532 nm, the wavelength for incoherent excitation of NV centers in diamond, the MFD was 9 μm × 9 μm and similar losses were found compared to 635 nm wavelength. By using an end-fire free-space coupling setup with an adjustable polarizer at the input, we found that the type II waveguide from Fig. 2(a) supported only the TM mode. Similar polarization-dependent behavior has been reported in type II waveguides in other crystals such as lithium niobate[22], KDP[35], and Ti:Sapphire[36] and can be attributed to the elliptical shape of the modification, which results in a different stress-induced refractive index distribution for TE and TM polarizations[22].

We further confirmed that 500-kHz offers a better regime for type II modification of diamond by lowering the repetition rate of the laser to 5 kHz. Over a wide range of laser processing parameters tested (200 nJ to 1 μJ pulse energy, track separations of 10 to 18 μm, scan speeds of 0.01 to 10 mm/s), we could not obtain waveguides with well-confined modes or reasonably low insertion loss. Waveguiding may be impeded by increased graphitization at lower repetition rates (Fig. 1(c)) resulting in higher absorption of the launched visible light, supported by absorption measurements on samples with tracks written over the entire 5 mm × 5 mm area (20 μm spacing between lines at 50 μm depth). As shown in Fig. 1(d), for visible wavelengths, the absorption is slightly higher for lower repetition rate processing. We also found that the optical bandgap was reduced for 5 kHz processing, which is expected when graphitic inclusions are present[37].

To achieve single mode guiding, we wrote a second vertically offset type II modification (Fig. 2(b)). This four-line modification supported a single mode at 635 nm wavelength with a MFD of 9 μm × 10 μm, an insertion loss of 19 dB and a coupling loss of 2.4 dB/facet. Similar to the case of the two-line type II waveguide, the four-line structure supported only the TM mode. Although each of the four lines shown in Fig. 2(b) was written with the same laser processing conditions, the vertical elongation of the lines increases dramatically with depth due to increased spherical aberration, which could be avoided by employing a spatial light modulator to impose a phase profile that is equal and opposite to the aberration introduced by the refractive index mismatch[38]. This would allow more flexible and precise 3D patterning of the barrier zones of the type II waveguide, enabling a more symmetric four-line structure or even an annular type modification for confining light.

μRaman measurements in diamond are sensitive to the presence of stress in the material. A shift of the diamond peak to higher (lower) wavenumber is associated with a compressive (tensile) stress. In Fig. 3(a) we present a map showing the peak shift near two laser written tracks. It can be seen that inside the guiding region the stress is mainly compressive, with a shift of about +1.5 cm$^{-1}$ compared to pristine diamond. A similar shift in the diamond peak is observed at the center of the four-line type II structure (Fig. 3(b)). However, previous studies have shown that compressive stress in diamond results in a decrease of the refractive index[39]. Therefore, the optical waveguiding we observe in the type II structures may be due to increased polarizability[40] or simply from a reduced refractive index in the modification lines, which serve as barriers for confining the optical mode.

Figure 4(b) plots the width of diamond Raman peak within the guiding region of the two-line type II waveguide. Within experimental uncertainty (1 cm$^{-1}$), the width of the crystal peak in the guiding region is the same as its value in pristine diamond (2.3 cm$^{-1}$). This is evidence that the crystalline structure in the guiding region is preserved, even if it is under compressive stress as shown in Fig. 3.

*NV center properties within waveguides*

The type II modification is beneficial for quantum information and magnetometry applications relying on NV centers, since the waveguide mode propagates in the undamaged region between the laser-inscribed lines. Photoluminescence (PL) characterization using a confocal microscope revealed a reduction in the intensity of the NV's zero phonon line (ZPL) transition within the modification lines (Fig. 5(a)). However, in the region between the modification lines where the optical mode propagates, the PL spectrum appears to be the same as that of pristine diamond. To give a quantitative estimate we compare the ratios of the integrated intensity of the NV$^-$ center ZPL to the 1$^{st}$ order Raman of the pristine diamond and waveguide spectra, which are 0.728 ± 0.006 and 0.733 ± 0.006, respectively. This suggests there is not a statistically significant change in the level of NV$^-$ centers within the waveguide region.

By launching 532-nm light into the waveguide and collecting the light using a spectrometer, we found a similar PL spectrum as that obtained by confocal microscopy (Fig. 5(b)). As the waveguide only supported the TM mode, the difference in the PL recorded from the TM and the TE configuration (shadowed in Fig. 5(b)) is the actual contribution of the guiding region to the PL, compared to the bulk material excited around the guide (Fig. 5(b) inset). These measurements provide strong evidence that the PL of the NV centers excited with the 532-nm laser inside the guiding region can be effectively collected and guided using the fabricated waveguide. Within the modification lines, a bright signal was observed between 730 nm and 800 nm, which is attributed to the GR1 color center (740 nm) associated with neutral vacancies[41] created by laser writing. It is possible that other vacancy complexes and interstitial defects are also induced causing spectral features similar to the radiation B Band of diamond which has undergone ion implantation and annealing[42].

Optically detected magnetic resonance (ODMR) characterization of the NV centers' electron spin within the buried diamond waveguides showed hyperfine structure (Fig. 6(a),(b)) with linewidths of $\Delta = 1.7 \pm 0.4$ MHz, comparable to those measured from pristine diamond ($\Delta = 1.8 \pm 0.3$ MHz) for the same measurement parameters. These linewidths provide a lower bound on the pure dephasing time $T_2^* = 1/(\pi\Delta)$ of the electron spin ensemble where $T_2^* = 0.2$ μs for the case of the NV centers in the waveguide. The linewidths recorded in these measurements are broader than those reported for similar grades of diamond (0.2-1 MHz[43]) though this could be caused by power broadening[44] and imperfect static field alignment[45]. Further, the lifetime of the excited state transition (11.0 ± 1.5 ns) was the same for NV centers within the waveguide and pristine diamond (Fig. 6(c)). These measurements provide evidence that the electronic structure of the NV centers was not severely affected by femtosecond laser inscription. Direct measurement of the spin coherence time of NV centers and the application of dynamical decoupling techniques will give a more complete picture of the effect of the laser writing process on the NV center and its coherence properties.

**Discussion**

The reduction in graphitization with MHz-repetition rates compared to kHz rates may be due to the higher temperatures driven by the higher pulse delivery rate. Jerng *et al.* observed that above growth temperatures of 1100°C, amorphous carbon was produced instead of nanocrystalline graphite films[46]. In agreement with characterization by μRaman spectroscopy, absorption and waveguide transmission, we have found that the electrical conductivity of lines written with the femtosecond laser is increased at lower repetition rates[47]. Future characterization including electro-chemical etching of the laser-written lines will seek to provide more insight into the modifications induced within the MHz repetition rate regime exploring other exposure parameters such as beam shaping with the SLM, pulse duration and wavelength.

Working at room temperature, magnetometry devices based on NV ensembles[48] would benefit from waveguides for efficient collection and routing of the fluorescence signal. Even more compelling is a diamond waveguide device with integrated Bragg reflectors at green and infrared wavelengths, to enable an integrated cavity enhanced magnetometer[49]. Bragg gratings waveguides have been demonstrated in glass by periodically modulating the intensity of the femtosecond laser pulse train during waveguide writing[50,51], and a similar approach will be applied to diamond. Diamond waveguides with Bragg reflection functionality would enable a significant enhancement of the magnetic-field sensitivity since the cavities increase the absorption path length by a factor proportional to their finesse[52].

Diamond has shown itself as an important material for quantum photonics, being a host to defect centers with atom-like properties having long-lived spin quantum states and well-defined optical transitions. The 3D photonics toolkit that we have developed demonstrating high quality NV properties while allowing easy interfacing to standard optical fibers may help diamond reach its full potential for quantum technologies.

**Methods**

The femtosecond second laser used for waveguide writing in diamond was a regeneratively amplified Yb:KGW system (Pharos, Light Conversion) with 230-fs pulse duration, 515-nm wavelength (frequency doubled), focused with a 1.25-NA oil immersion lens (RMS100X-O 100× Olympus Plan Achromat Oil Immersion Objective, 100× oil immersion, Olympus). Employing such a high NA allows for a smaller focal volume, to minimize the writing power and avoid self-focusing (critical power 2 MW at 515 nm), which lead to vertically elongated waveguide cross sections and non-reproducible results.

The polarization of the incident laser was perpendicular to the scan direction. The repetition rate of the laser was variable from 1 MHz to single pulse. Computer-controlled, 3-axis motion stages (ABL-1000, Aerotech) interfaced by CAD-based software (ScaBase, Altechna) with an integrated acousto-optic modulator (AOM) were used to translate the sample relative to the laser to form the desired photonic structures.

Polished 5 mm × 5 mm × 0.5 mm synthetic single-crystal diamond samples (type II, optical grade with nitrogen impurities 100 ppb) were acquired from MB Optics. Laser-inscribed structures were characterized for their morphology using white-light optical microscopy in transmission mode with 10× and 40× magnification objectives (Eclipse ME600, Nikon). We also fabricated waveguides in electronic grade single-crystal diamond (nitrogen impurities 5ppb, dimensions 3 mm × 3 mm × 0.2 mm) from MB optics which revealed similar propagation losses and mode profiles.

For waveguide transmission measurements, high resolution 3-axis manual positioners (Nanomax MAX313D, Thorlabs) were used. The four-axis central waveguide manipulator (MicroBlock MBT401D, Thorlabs) enabled transverse displacement between sets of diamond waveguides. Light sources at 808 nm (S1FC808, Thorlabs), 635 nm (TLS001-635, Thorlabs) and 532 nm (4301-010, Uniphase) were coupled to the waveguides using the appropriate Thorlabs single-mode fibers for each of the visible wavelengths (460HP for 532 nm, SM600 for 635 nm, 780HP for 808 nm). To test the polarization dependence of the waveguide transmission, free-space coupling was used with 10× (5721-H-B, Newport) and 60× (5721-H-B, Newport) lenses at the input and output, respectively. At the output, light was coupled to an optical power meter (818-SL, Newport) to measure the power transmitted through the waveguide. To measure the near-field waveguide mode profile, a 60× asphere (5721-H-B, Newport) was used to image the light to a beam profiler (SP620U, Spiricon).

Micro-Raman spectra were recorded using a Labram Aramis Jobin Yvon Horiba microRaman system with a DPSS laser source of 532 nm and equipped with a confocal microscope and an air-cooled CCD. A 50× (100×) objective was used to focus the laser on the sample as well as to collect the Raman signal, with a spatial resolution of about 1 micron. A wavenumber accuracy of about 1 cm$^{-1}$ can be achieved with a 1800 line/mm grating.

To observe the characteristic fluorescence from the negatively charged NV center at 637 nm (and the phonon sidebands), 532-nm light from a 1-W frequency-doubled Nd:YVO$_4$ laser (Verdi, Coherent) was coupled with and collected with the free space optics described above and detected with a spectrometer (Ocean Optics model HR2000). A notch filter in the green was used to attenuate the pump wavelength.

For confocal photoluminescence measurements, nitrogen-vacancy defects were excited with a DPPS 532-nm laser (CL532-500-L, CrystaLaser) focused on to the sample with a 0.55 NA objective (100X Plan Apo SL Infinity Corrected Objective, Mitutoyo). Photoluminescence was collected through the same objective, filtered from the excitation light using a dichroic beamsplitter (ZT 532 RDC, Chroma) and long-pass filters (ET 555 LP Chroma, FELH 0650 Thorlabs) and focused into a single mode fibre which provided the confocal aperture. Photon counting of the filtered light was performed using an avalanche photodiode (SPQR-14, Perkin-Elmer). Optically detected magnetic resonance measurements were performed by monitoring the fluorescence rate while scanning the frequency of a microwave field driven through a 20 μm copper wire on the surface of the sample generated using a commercial microwave source (Agilent ESG E4433B) amplified by a high power broad band amplifier (Minicircuits ZHL-16W-43+). A static magnetic field was applied using a permanent magnet. The field was ~90 G for the measurements in Fig. 6(a) and (b). Lifetime measurements were performed using a supercontinuum source for excitation filtered at 532 nm (Fianium Whitelase WL-SC400-4), and a time correlated single photon counting board (Timeharp 260).

**Acknowledgements**

This work has been supported by the FP7 DiamondFab CONCERT Japan project, DIAMANTE MIUR-SIR grant, and FemtoDiamante Cariplo ERC reinforcement grant. We thank Michael Burek, Patrick Salter, Paolo Olivero and Federico Bosia for helpful scientific discussions. We gratefully thank Ursula Klabbers of MB Optics (Velp, Netherlands) for assistance in post-polishing of the diamond samples.

**Author Contributions**

S. M. E. and P. E. B. conceived the idea of forming waveguides in the bulk of diamond. V. B., B. S., L. C. and S. M. E. performed femtosecond laser writing experiments in diamond. J. P. H., P. B. and S. M. E. characterized the properties of NV centers within the optical waveguides. B. S. and V. B. performed mode profile and insertion loss measurements of the optical waveguides. B. S., V. B., T. T. F., G. G. and R. O. characterized the photoluminescence collected at the output of waveguides. B. S., A. C., M. S., O. J., Y. S., K. M., R. R. and M. F. characterized the structure of femtosecond laser written lines using μRaman spectroscopy. S. L. applied theoretical calculations to estimate the propagation loss due to graphitic lines. All authors discussed the experimental implementation and results and contributed to writing the paper.

**Competing financial interests**

The authors declare no competing financial interests.

*Published in* **Scientific Reports 6, Article number: 35566 (2016) doi:10.1038/srep35566**## References

1. Childress, L., Taylor, J., Sørensen, A. S. & Lukin, M. D. Fault-tolerant quantum repeaters with minimal physical resources and implementations based on single-photon emitters. *Physical Review A* **72**, 052330 (2005).
2. Childress, L., Walsworth, R. & Lukin, M. Atom-like crystal defects. *Physics Today* **67**, 38 (2014).
3. Balasubramanian, G. *et al.* Ultralong spin coherence time in isotopically engineered diamond. *Nature materials* **8**, 383-387 (2009).
4. Schirhagl, R., Chang, K., Loretz, M. & Degen, C. L. Nitrogen-vacancy centers in diamond: nanoscale sensors for physics and biology. *Annual review of physical chemistry* **65**, 83-105 (2014).
5. Hensen, B. *et al.* Loophole-free Bell inequality violation using electron spins separated by 1.3 kilometres. *Nature* **526**, 682-686 (2015).
6. Aharonovich, I., Greentree, A. D. & Prawer, S. Diamond photonics. *Nature Photonics* **5**, 397-405 (2011).
7. Kimble, H. J. The quantum internet. *Nature* **453**, 1023-1030 (2008).
8. Doerr, C. R. & Okamoto, K. Advances in silica planar lightwave circuits. *Journal of Lightwave Technology* **24**, 4763-4789 (2006).
9. Soref, R. The past, present, and future of silicon photonics. *Selected Topics in Quantum Electronics, IEEE Journal of* **12**, 1678-1687 (2006).
10. Barclay, P. E., Fu, K.-M. C., Santori, C., Faraon, A. & Beausoleil, R. G. Hybrid nanocavity resonant enhancement of color center emission in diamond. *Physical Review X* **1**, 011007 (2011).
11. Fu, K.-M. *et al.* Coupling of nitrogen-vacancy centers in diamond to a GaP waveguide. *Applied Physics Letters* **93**, 234107 (2008).
12. Olivero, P. *et al.* Ion-Beam-Assisted Lift-Off Technique for Three-Dimensional Micromachining of Freestanding Single-Crystal Diamond. *Advanced Materials* **17**, 2427-2430 (2005).
13. Faraon, A., Santori, C., Huang, Z., Acosta, V. M. & Beausoleil, R. G. Coupling of nitrogen-vacancy centers to photonic crystal cavities in monocrystalline diamond. *Physical review letters* **109**, 033604 (2012).
14. Burek, M. J. *et al.* Free-standing mechanical and photonic nanostructures in single-crystal diamond. *Nano letters* **12**, 6084-6089 (2012).
15. Faraon, A., Barclay, P. E., Santori, C., Fu, K.-M. C. & Beausoleil, R. G. Resonant enhancement of the zero-phonon emission from a colour centre in a diamond cavity. *Nature Photonics* **5**, 301-305 (2011).
16. Hausmann, B. *et al.* Coupling of NV centers to photonic crystal nanobeams in diamond. *Nano letters* **13**, 5791-5796 (2013).
17. Burek, M. J. *et al.* High quality-factor optical nanocavities in bulk single-crystal diamond. *Nature communications* **5** (2014).
18. Khanaliloo, B. *et al.* Single-Crystal Diamond Nanobeam Waveguide Optomechanics. *Physical Review X* **5**, 041051 (2015).
19. Lagomarsino, S. *et al.* Evidence of light guiding in ion-implanted diamond. *Physical review letters* **105**, 233903 (2010).
16

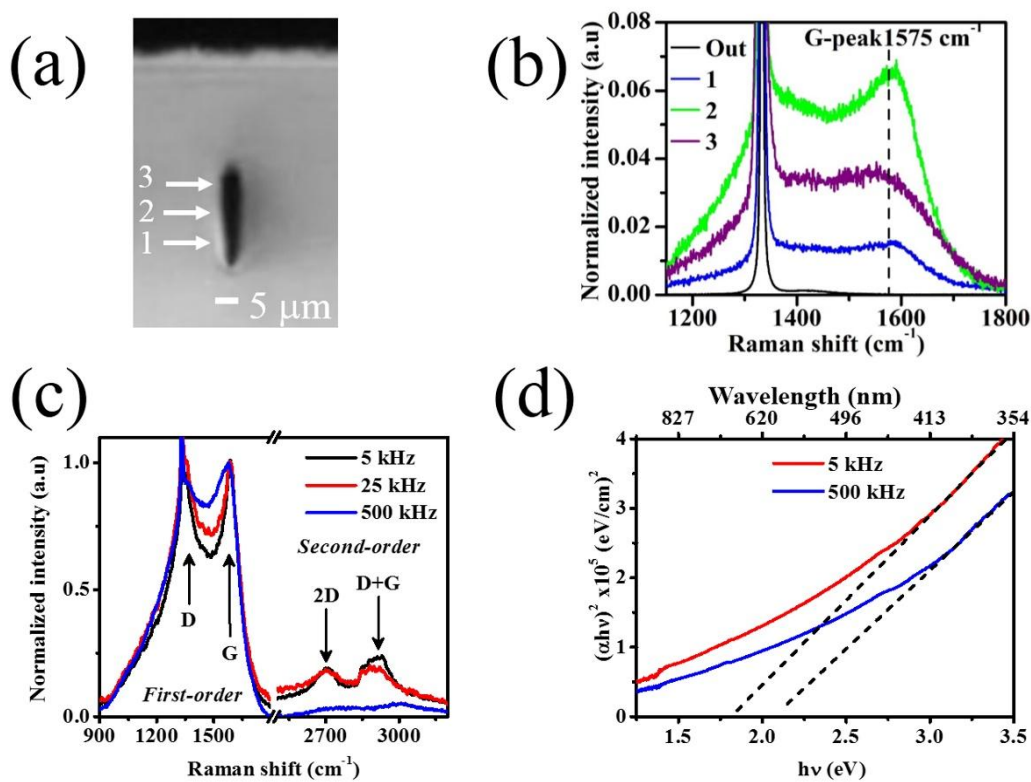

**Figure 1. Suppression of graphite in femtosecond laser induced modification at 500 kHz repetition rate**. (a) Transverse optical microscope image of single laser-induced track written with 500-kHz repetition rate, 50-mW average power and 0.5-mm/s scan speed. (b) µRaman spectra (532-nm excitation wavelength) at four different vertical positions inside the modification. 'Out' refers to a spectrum taken outside the track. The spectra have been normalized to the diamond peak to show the change in the relative intensity of the G-peak inside the structure. (c) µRaman spectra (normalized to the G-peak) in the center of modification tracks at repetition rates of 5 kHz, 25 kHz and 500 kHz, with pulse energy held constant (800 nJ) to produce a similar size modification at each repetition rate. (d) Tauc plot for diamond with tracks written over the entire sample at 50-µm depth and 20-µm line separation for 5-kHz and 500-

kHz repetition rates. It is considered that the absorption in the visible region is only due to the modification tracks, with the rest of the sample being transparent.

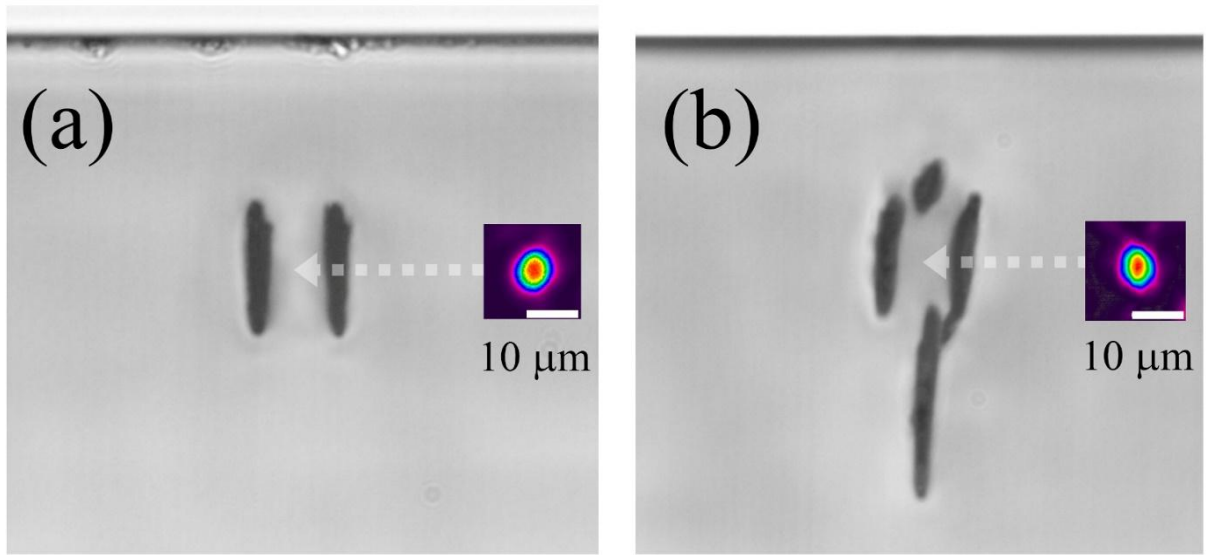

**Figure 2. Type II buried optical waveguides in diamond.** Transverse microscope view of type II waveguide in diamond along with near field mode profile ($\lambda$ = 635 nm). An arrow indicates the position of the mode. (a) Pair of lines, horizontally separated by 13 μm. Modes could be coupled into three different vertical positions with the lowest loss mode shown (MFD 10 μm × 11 μm). (b) Pair of lines, horizontally separated by 13 μm along with second pair of lines for vertical confinement. Only a single mode could be coupled to the four-line modification (MFD 9 μm × 10 μm). All tracks were written with 50 mW, 0.5 mm/s at 500 kHz with deeper tracks more elongated due to increased spherical aberration.

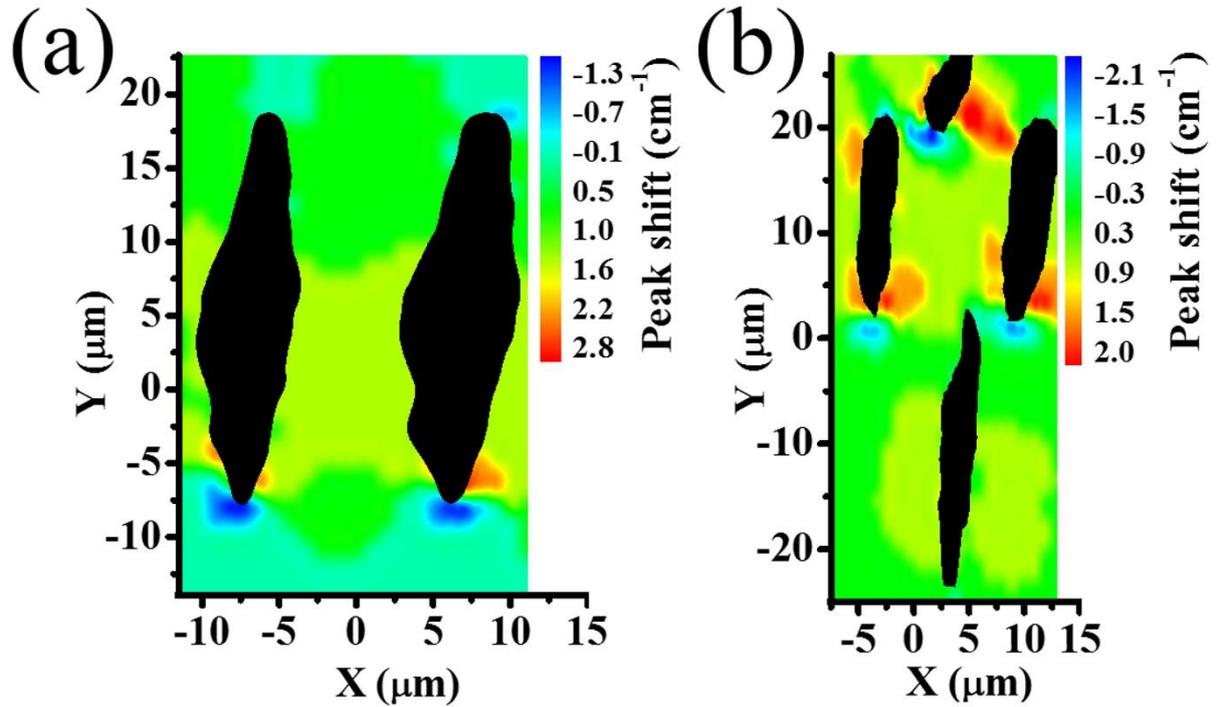

**Figure 3. Peak shift map of type II waveguides using µRaman spectroscopy.** Spatial map of frequency shift of diamond Raman peak with respect to bulk for (a) two-line and (b) four-line modification, with the same parameters as those in Figure 2. The modification tracks are shown as black.

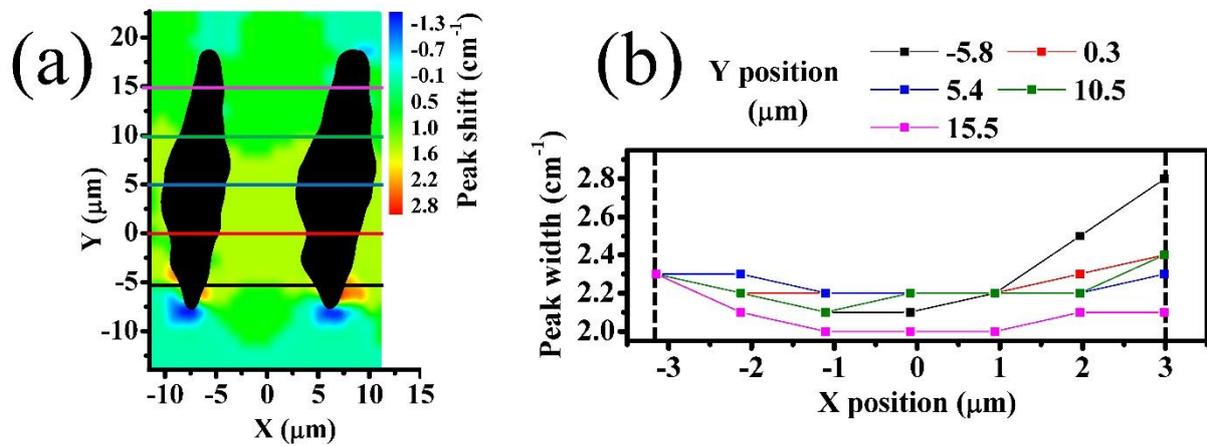

**Figure 4. Width of diamond Raman peak within waveguiding region**. (a) Peak shift map from Fig. 3(a) with colored lines corresponding to the line profiles (b) of the width of diamond Raman peak inside the guiding region. Dotted lines mark the approximate position of the edges of the tracks.

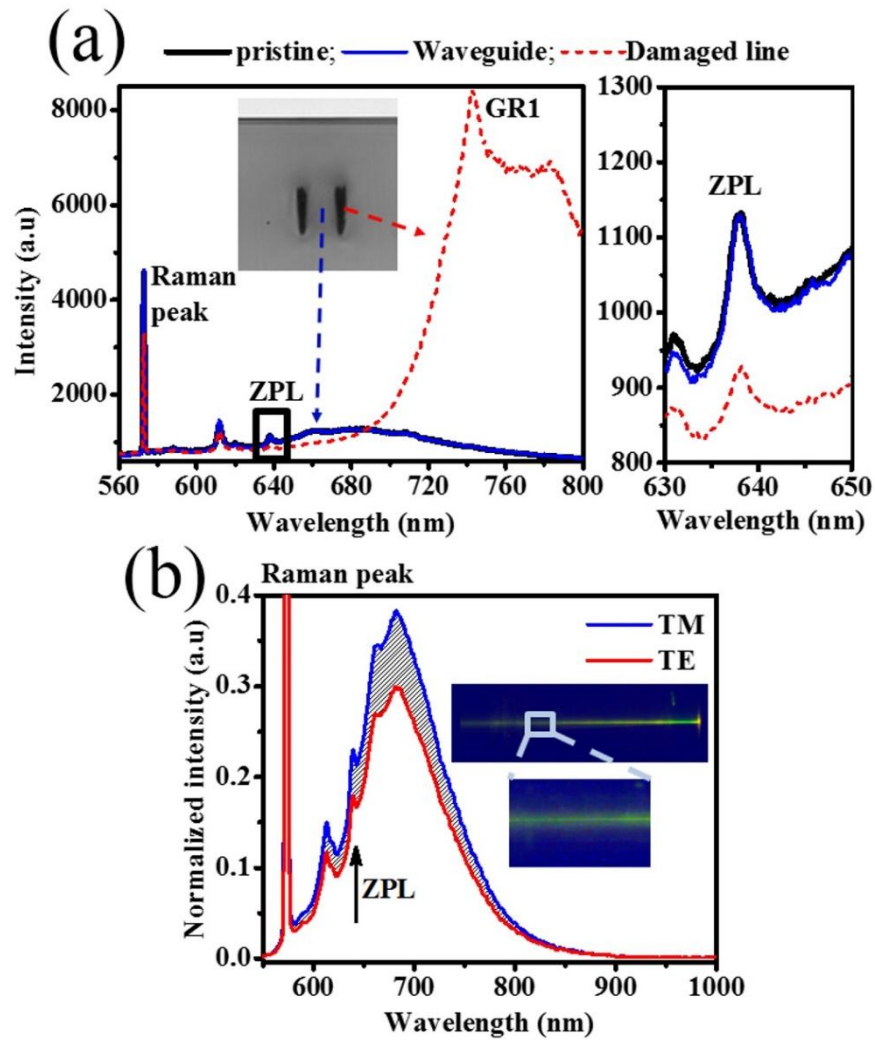

**Figure 5. Photoluminescence spectra in waveguide.** (a) Photoluminescence spectra within the laser-written lines, between the modification lines in the waveguiding region and in pristine diamond acquired by confocal microscopy (excitation wavelength 532 nm). A cross sectional microscope image of the type II waveguide is shown in the inset. A zoomed in view of the ZPL spectrum (indicated with a black square) is presented on the right, showing that it remains unchanged in the guiding region compared to pristine diamond. (b) Photoluminescence detected with spectrometer at output of the waveguide when light was coupled in using free space optics (TM and TE configurations are presented). The inset shows

an overhead microscope image of the fluorescence streak when the 532-nm light was coupled to the waveguide.

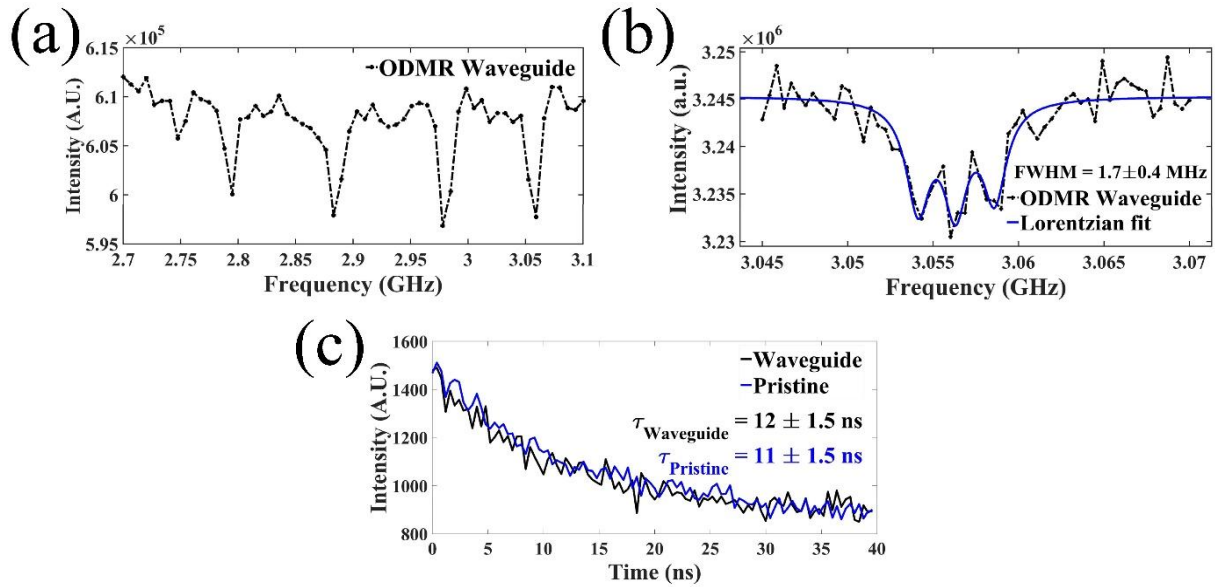

**Figure 6. Preservation of hyperfine structure of ground state and excited state lifetime.** (a) Coarse ODMR scan within the waveguide where the electron spin transitions from $m_s = 0 \rightarrow m_s = \pm 1$ are indicated by a drop in fluorescence intensity. Several families of transitions are visible because of Zeeman splitting caused between a static magnetic field (~90 G) and the four possible orientations of the NV center within the diamond lattice. (b) Finer ODMR scan of one of the electron spin transitions indicated in (a). This shows the hyperfine structure the electron spin coupled the NV centers' $^{14}$N nuclear spin. The three transitions correspond to $m_s = 0 \rightarrow m_s = +1$ in the electron spin, with the hyperfine coupling to the nuclear spin splitting the $m_s = +1$ state into 3 for the three projections of the $S = 1$ nuclear spin. The transitions were fit to Lorentzians yielding a FWHM of $1.7 \pm 0.4$ MHz. (c) Lifetime measurement within waveguide (blue curve) and in pristine diamond (black curve) of the excited state transition fit to $e^{-t/\tau}$.